\documentclass[aps,pra,twocolumn]{revtex4}

\usepackage{color}
\usepackage{mathtools}
\usepackage{graphicx}
\usepackage{ulem}
\usepackage{float}

\begin{document}

\title{Interplay of photons and charge carriers in thin-film devices}

\author{Pyry Kivisaari, Mikko Partanen, Toufik Sadi, and Jani Oksanen}
\affiliation{Engineered Nanosystems Group, Aalto University, P.O. Box 13500, FI-00076 Aalto, Finland}

\begin{abstract}
Thin films are gaining ground in photonics and optoelectronics, promising improvements in their efficiency and functionality as well as decreased material usage as compared to bulk technologies. However, proliferation of thin films would benefit not only from continuous improvements in their fabrication, but also from a unified and accurate theoretical framework of the interplay of photons and charge carriers. In particular, such a framework would need to account quantitatively and self-consistently for photon recycling and interference effects. To this end, here we combine the drift-diffusion formalism of charge carrier dynamics and the fluctuational electrodynamics of photon transport self-consistently using the recently introduced interference-extended radiative transfer equations. The resulting equation system can be solved numerically using standard simulation tools, and as an example, here we apply it to study well-known GaAs thin-film solar cells. In addition to obtaining the expected device characteristics, we analyze the underlying complex photon transport and recombination-generation processes, demonstrating the physical insight provided for unevenly excited structures through the direct and self-consistent description of photons and charge carriers. The methodology proposed in this work is general and can be used to obtain accurate physical insight into a wide range of planar optoelectronic devices, of which the thin-film single-junction solar cells studied here are only one example.
\end{abstract}


\maketitle

\section{Introduction}

Photonics and optoelectronics have irrevocably transformed several areas of society, including energy production through solar cells, general lighting through light-emitting diodes, and high-speed Internet through global optical fibre networks \cite{green_2016,cho_2017,bayvel_2016}. As these technologies mature and become established, new structures and device concepts are emerging both to increase the efficiency of existing technologies and to create new application areas. One such emerging technology is thin and ultra-thin film devices incorporating 0.1-10 $\mu$m thick semiconductor layers that enable, e.g., reusing the growth substrate several times and utilizing optical cavity effects \cite{cappelluti_2018,sai_2019,eerden_2019,massiot_2014,metaferia_2019,schermer_2005}. However, taking full advantage of these possibilities would still greatly benefit from improved theoretical understanding, in particular in self-consistently accounting for the interplay between photons and charge carriers as well as in combining wave optical resonance effects with the quantum optical loss mechanisms. Having accurate, generally applicable and insightful modeling frameworks of these effects would facilitate assessing e.g. the overall role of photon recycling in high-efficiency solar cells. To that end, in this paper we propose an accurate and fully self-consistent electro-optical modeling tool for planar thin-film optoelectronic devices. The model is directly derived from fluctuational electrodynamics, and thanks to the coupling with charge-carrier dynamics, it self-consistently includes photon recycling and all the pertaining interference effects also for the case of spatially uneven excitation.

\begin{figure}
\centering
\includegraphics[width=0.8\columnwidth]{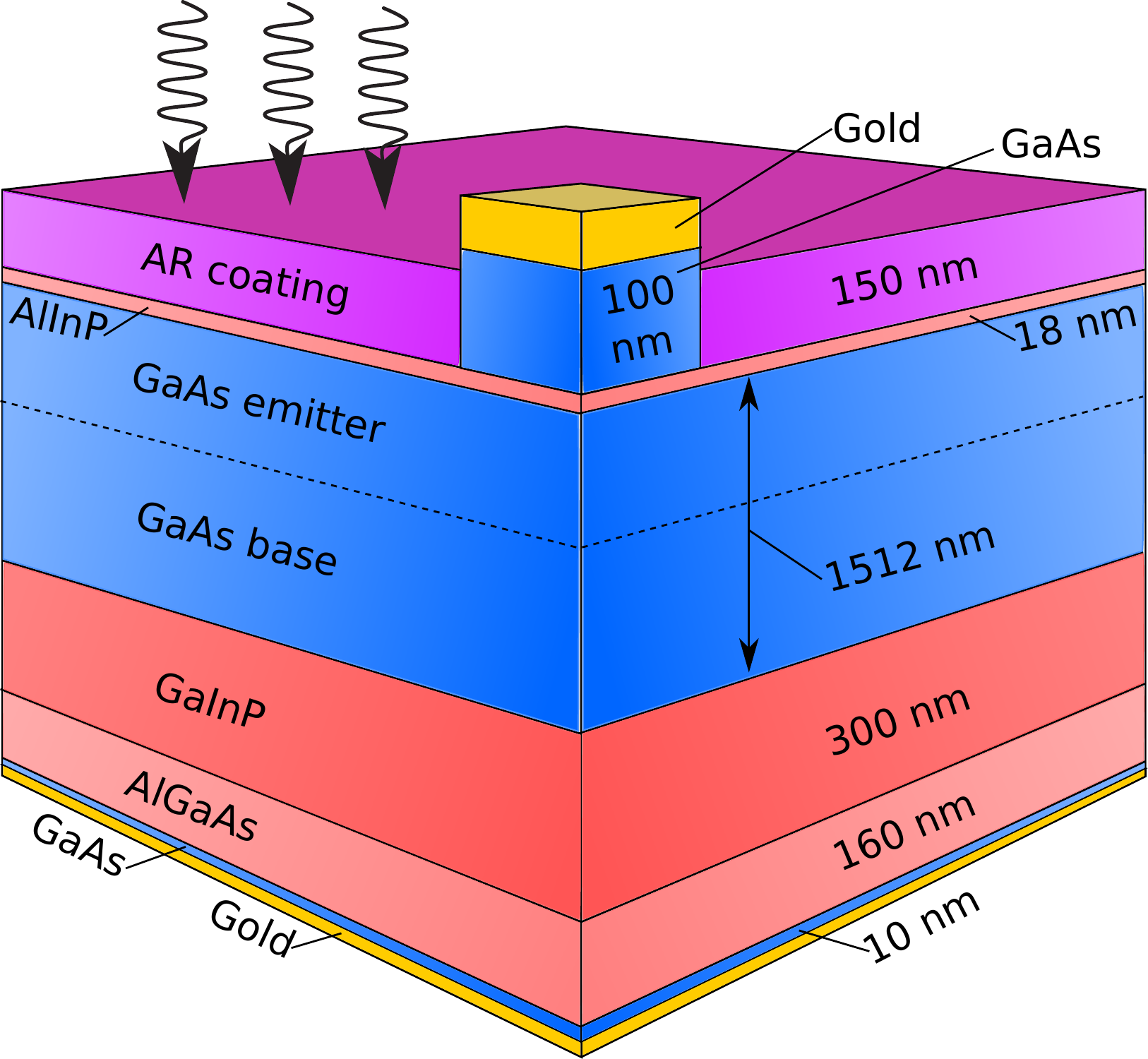}
\caption{Thin-film solar cell structure studied in this paper using the proposed interference radiative transfer--drift-diffusion (IRTDD) model. The GaAs/gold contact grid is not included in the optical calculation, which presently solves Maxwell's equations accurately for planar layer structures.}
\label{fig:structure}
\end{figure}

Due to its general importance, electro-optical simulation of resonant optoelectronic devices has been a topic of research already for a few decades. Selected highlights of the state of the art from the recent literature include using transfer matrices to calculate coupling matrices in order to account for photon recycling and luminescent coupling in drift-diffusion (DD) simulations \cite{wilkins_2015}, and calculating position-dependent dipole emission in a layer structure to scale the local radiative recombination coefficient and create a photon redistribution matrix to be used in the DD simulation \cite{walker_2015}. Summarizing from these, one typically starts with optical solution methods, such as transfer or scattering matrices or finite-difference time-domain simulations, to create coupling matrices or other constructions relating emission and absorption locations with each other. These are then used in the carrier dynamics simulation to account for electro-optical effects such as photon recycling. However, previous works have not presented a full analysis of how the combination of illumination and applied bias affects the spectral, directional and resonant effects in the generation-recombination profiles of the structures. The aim of this paper is to facilitate such analyses and complement existing modeling frameworks with a direct modeling tool where the optical and electrical properties are solved simultaneously, either using a direct or iterative solver. This enables a direct and accurate calculation of the position-dependent recombination-generation rate in unevenly excited planar structures, fully accounting for the rigorously calculated 3D optical mode structure.

The full electro-optical device modeling tool proposed in this paper is implemented by combining the well-known DD model with fluctuational electrodynamics (FED) using the recently introduced interference-extended radiative transfer (IRT) model \cite{partanen_2017_sr}. Unlike the conventional radiative transfer equation \cite{rte}, the IRT model includes all wave-optical and photon recycling effects self-consistently as determined by both the external optical field boundary conditions and the applied bias. On the other hand, the underlying FED is a natural choice for the goals of this paper, as it combines Maxwell's equations directly with stochastic sources using the dyadic Green's functions. FED therefore enables rigorously calculating the interaction between electron-hole distributions following position-dependent quasi-Fermi levels and the full electromagnetic mode structure by building on our comprehensive recent work on the fundamental aspects (e.g. Refs. \citenum{partanen_2017_sr,partanen_2017,kivisaari_2018}). It is also noteworthy that the IRT represents a rigorous solution of Maxwell's equations in the planar case, and it thus includes all the electromagnetic information that is alternatively obtained e.g. using transfer matrices. To demonstrate the model with a well-known device structure, here we apply it to simulate the thin-film solar cell structure studied in Ref. \citenum{steiner_2013}. We show that the model reveals delicate optical energy transfer processes that are really feasible to study only if one has access to a fully self-consistent model, such as the one proposed here. By doing this, we wish to illustrate the insight that can be gained with the proposed interference radiative transfer--drift-diffusion (IRTDD) model, and how it can be used to gain detailed physical understanding of essentially any planar optoelectronic device.

\section{Theory} \label{sec:theory}

In this Section, we describe the IRTDD model used to carry out fully self-consistent electro-optical simulations of thin-film optoelectronic devices. Present version of the model combines the quantized fluctuational electrodynamics (QFED) framework in planar layer structures directly with charge-carrier dynamics as governed by the DD equations, but we expect that a similar IRTDD model could also be derived from the classical fluctuational electrodynamics. We start by writing the IRT model in a somewhat simplified form for a given polarization as \cite{partanen_2017_sr}
\begin{equation}
\begin{cases}
 \frac{d}{dz}\phi(z,K,\omega) & = -\alpha(z,K,\omega)[\phi(z,K,\omega)-\eta(z,\omega)] \\
 & \hspace{0.4cm}+\beta(z,K,\omega)[\phi(z,K,-\omega)-\eta(z,\omega)] \vspace{0.05in} \\
S(z,K,\omega) & = \frac{1}{2}\hbar\omega v(z,K,\omega)\rho(z,K,\omega) \\
 & \hspace{0.4cm}\times \left[\phi(z,K,\omega)-\phi(z,K,-\omega)\right].
\end{cases}
\label{eq:irte}
\end{equation}
The first part of Eq.~(\ref{eq:irte}) is the interference-extended radiative transfer equation for the photon number $\phi$ of a mode described by its in-plane wave number $K$, angular frequency $\omega$ (with $-\omega$ here denoting the modes propagating towards the negative direction, i.e. downwards in Fig.~\ref{fig:structure}), and position coordinate in the normal direction $z$. Moreover, $\alpha$ and $\beta$ are the damping and scattering coefficients formulated with help of the dyadic Green's functions as in Ref. \citenum{partanen_2017_sr}, and they represent a rigorous solution to Maxwell's equations in the planar layered geometry. Finally, $\eta(z,\omega)$ is the Bose-Einstein distribution function $\eta=1/\left(e^{(\hbar\omega-\Delta E_F)/(k_BT)}-1\right)$ that represents the equilibrium photon distribution and acts as a local radiative source term. Here, the photon chemical potential is assumed equal to the local quasi-Fermi level separation $\Delta E_F$ following the arguments presented in Ref. \cite{wurfel_1982}.

The second part of Eq.~(\ref{eq:irte}) relates the photon number with the spectral radiance $S$ in the direction determined by $K$ and $\omega$. Here, $\rho$ is the local density of states as defined in Refs. \citenum{partanen_2017_sr,partanen_2017}, and $v$ is the generalized speed of light defined in Appendix \ref{sec:appendixa}. As the quantities $v$, $\rho$, $\alpha$ and $\beta$ are derived from fluctuational electrodynamics, they capture all wave-optical effects such as internal reflections, constructive and destructive interferences and emission enhancement and suppression, and they are geometry dependent. The local generation-recombination rate can be calculated with help of the derivative of $S(z,K,\omega)$ as
\begin{equation}
R_{r}=\displaystyle\sum_{pol.}\int_0^{\infty}d\omega\int_0^{\infty}dK\frac{dS}{dz}\frac{1}{\hbar\omega}2\pi K,
\label{eq:Rrad}
\end{equation}
where the sum is taken over the two orthogonal polarizations of light. The derivative of $S$ is trivial to calculate using the chain rule and the equations provided in Ref. \citenum{partanen_2017_sr}. Through the local densities of states included in $S$ and the 2D direction integral accounted for by $2\pi K$, $R_r$ includes the contributions of all polar angles and symmetric in-plane directions. Furthermore, we have recently introduced a straightforward way to calculate the dyadic Green's function, the local and nonlocal optical densities of states, and the $\alpha$ and $\beta$ coefficients for arbitrary planar structures with help of optical admittances \cite{kivisaari_2018}, and it is used also in this paper.

Direction-dependent incoming solar radiation can be readily included as a boundary condition for $\phi$, in which case it is also included in the local generation-recombination rate $R_r$. However, if one wishes to project the incoming sunlight fully to the normally incident mode ($K=0$), it can be more practical to calculate a separate solar generation rate $G_s$ with conventional methods or by using an additional IRT equation for the absorption of sunlight (corresponding to e.g. a transfer matrix solution). In that case, $R_r$ represents the recombination-generation of only those photons that are emitted inside the structure, and $G_s$ accounts only for the absorption of incident sunlight. Here, we write an additional IRT equation for a solar photon number $\phi_s(z,K=0,\omega)$, whose incoming boundary value $\phi_{s,inc}$ is selected such that the $S_{s,inc}$ calculated from $\phi_{s,inc}$ through Eq. (\ref{eq:irte}) is equal to the incoming spectral intensity of sunlight. As the emission inside the structure is fully accounted for already in $R_r$, this additional IRT equation for sunlight should not include the $\eta$ terms anymore. The solar generation rate $G_s$ can then be calculated from $\phi_s$ similarly as in Eq. (\ref{eq:Rrad}) by omitting the integration over $2\pi KdK$ and multiplying by $-1$.

The IRT can be readily coupled with the DD model of carrier transport, given by
\begin{equation}
\begin{array}{c}
\frac{d}{dz}\left(-\varepsilon\frac{d}{dz} U\right)=e\left(p-n+N\right),\\
\frac{d}{dz}J_{n}=\frac{d}{dz}\left(\mu_{n}n\frac{d}{dz} E_{Fn}\right)=e(R_r-G_s+R_{nr}),\\
\frac{d}{dz}J_{p}=\frac{d}{dz}\left(\mu_{p}p\frac{d}{dz} E_{Fp}\right)=-e(R_r-G_s+R_{nr}),
\end{array}\label{eq:DDyhtalot}
\end{equation}
where $\varepsilon$ is the static permittivity, $U$ is the electrostatic potential, $e$ is the elementary charge, $n,p$ are the electron and hole densities, $N$ is the ionized doping density (positive for donors and negative for acceptors), $J_n,J_p$ are the electron and hole current densities, $\mu_n,\mu_p$ are the electron and hole mobilities $E_{Fn},E_{Fp}$ are the conduction and valence band quasi-Fermi levels such that $\Delta E_F=E_{Fn}-E_{Fp}$, and $R_{nr}$ is the nonradiative recombination rate calculated as a sum of Shockley-Read-Hall and Auger recombination as in Ref. \citenum{heikkila_2009}. Radiative recombination-generation rate $R_r$ is calculated from the photon numbers using Eq.~(\ref{eq:Rrad}), and therefore the resulting IRTDD model constructed by Eqs. (\ref{eq:irte}) and (\ref{eq:DDyhtalot}) solves photon and charge carrier dynamics self-consistently. Moreover, the result is an accurate representation of the FED, connecting thermalized carrier distributions with the rigorously calculated optical mode structure. As an accurate and fully self-consistent combination of fluctuational electrodynamics and drift-diffusion dynamics, the IRTDD model provides new physical insight particularly into the optical energy transfer in unevenly excited structures, as demonstrated through the results in Section \ref{sec:results}.

The boundary conditions and integration limits required for the IRTDD model can be chosen as follows. For each photon energy, the IRT simulation is run for an equally-spaced set of $K$ between zero and suitably chosen $K>Nk_0$, where $N$ is the largest refractive index present in the semiconductor layers. According to the results of this paper, spectral radiances and generation-recombination rates all go to zero at $K>Nk_0$ as expected, indicating that the simulations span all relevant optical modes. For the DD model, the boundary conditions for the electrostatic potential and quasi-Fermi levels are determined by the built-in potential and applied bias in the conventional way. More specifically, $U$ is equal to 0 at the n-contact and $V_a-V_0$ at the p-contact, where $V_0$ is the built-in potential and $V_a$ is the applied bias. The quasi-Fermi levels of electrons and holes are equal with each other at the contacts ($=0$ at the n-contact and $qV_a$ at the p-contact), corresponding to infinitely fast contact recombination of minority carriers that reach the opposite contact. For incoming solar light, we use the spectral intensity according to the typical ASTM G173 global reference spectrum \cite{astm_2012} to enable direct comparison with device characteristics reported elsewhere. As written earlier, the incoming solar intensity is fully projected into the normally incident modes for simplicity. The incoming boundary value for $\phi_{s}$ is given by $\phi_{s,inc}=2S_{s,inc}/(\hbar\omega v_0\rho_0)$, where $S_{s,inc}$ is the ASTM G173 intensity written in W/($\frac{1}{\text{s}}$m$^2$), and $v_0$ and $\rho_0$ are the speed of light and local density of states calculated at $K=0$ on top of the structure. Then, with the external optical field already accounted for in $\phi_s$, the incoming values of $\phi$ are all set to zero. Both the DD and IRT equations are presently solved with the finite element method (FEM). Equation (\ref{eq:irte}) is solved for each $\omega$ and $K$, and the solution is iterated with Eq.~(\ref{eq:DDyhtalot}) through the use of Eq.~(\ref{eq:Rrad}). Convergence is reached for each applied bias by monitoring the change in the total current. Parallelizing the calculation of each $K$ and otherwise optimizing the solution procedure should significantly reduce the time required for a single bias point from the present approx. 15 minutes on a normal desktop computer.

\section{Results \& discussion} \label{sec:results}

To demonstrate the IRTDD model with a widely known and well-studied example device structure, we simulate a thin-film GaAs solar cell structure similar to the one studied in Ref. \citenum{steiner_2013} and shown schematically in Fig.~\ref{fig:structure}. This device is selected both to confirm that the IRTDD model reproduces the expected device characteristics and to demonstrate the new microscopic insight it provides even for a conventional thin-film solar cell. The device operation is simulated both under dark current conditions and under direct sunlight hitting the sample at normal incidence. The material parameters and the dielectric functions used in the simulations are specified in Appendix \ref{sec:appendixb}. We start by presenting the overall device characteristics resulting from the IRTDD simulation and proceed to studying the internal photon transport and recombination-generation properties under illumination and in dark current conditions. For the present purposes, we limit the geometry to be fully planar and homogeneous in the lateral direction. Most importantly, here this means that similar to Ref. \citenum{walker_2015}, the sparse top contact grid is not included in the optics calculation. It is, however, expected that the results and the model can be generalized for higher dimensional geometries or light scattering surfaces by separately accounting for lateral effects and extending the boundary conditions.

\subsection{Overall device characteristics}

\begin{figure}
\centering
\includegraphics[width=\columnwidth]{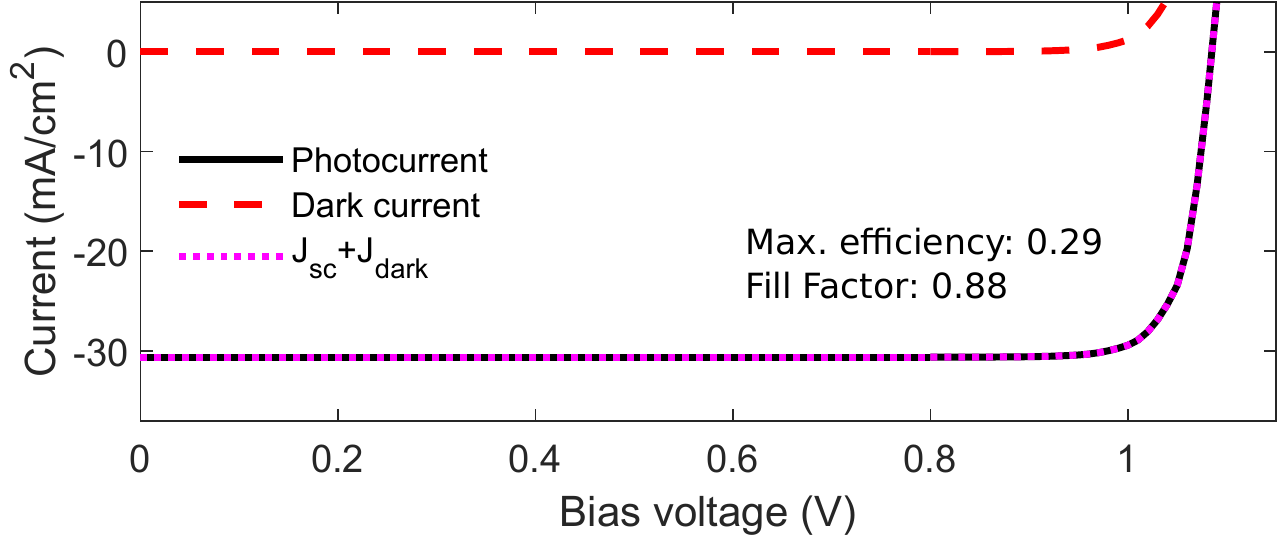}
\caption{Current-voltage characteristics of the structure simulated with the IRTDD model in dark and under normally incident sunlight corresponding to the ASTM G173 global reference spectrum. The photocurrent according to the superposition principle $J_{sc}+J_{dark}$ (sum of short-circuit and dark current densities) is also plotted.}
\label{fig:iv_nrel}
\end{figure}

Beginning with the overall device operation, Fig.~\ref{fig:iv_nrel} shows the current-voltage characteristics from the full IRTDD simulation both under illumination and under dark current conditions. The photocurrent curve shows the expected high-efficiency GaAs solar cell characteristics with a short-circuit current of 30.61 mA/cm$^2$, open-circuit voltage of 1.09 V, fill factor of 0.88 and maximum power efficiency of 0.29. All these values are close to the experimental and theoretical values reported in Ref. \citenum{steiner_2013}, with the main differences likely to result from slightly different simulation or material parameters, and possibly also from optical interference and carrier transport effects that were not included in the calculations of Ref. \citenum{steiner_2013}. Figure \ref{fig:iv_nrel} also includes the plot of $J_{sc}+J_{dark}(V_a)$, where $V_a$ is the applied bias, and $J_{sc}$ and $J_{dark}$ are the short-circuit and dark currents.

\subsection{Characteristics under illumination} \label{sec:illumination}

\begin{figure}
\centering
\includegraphics[width=\columnwidth]{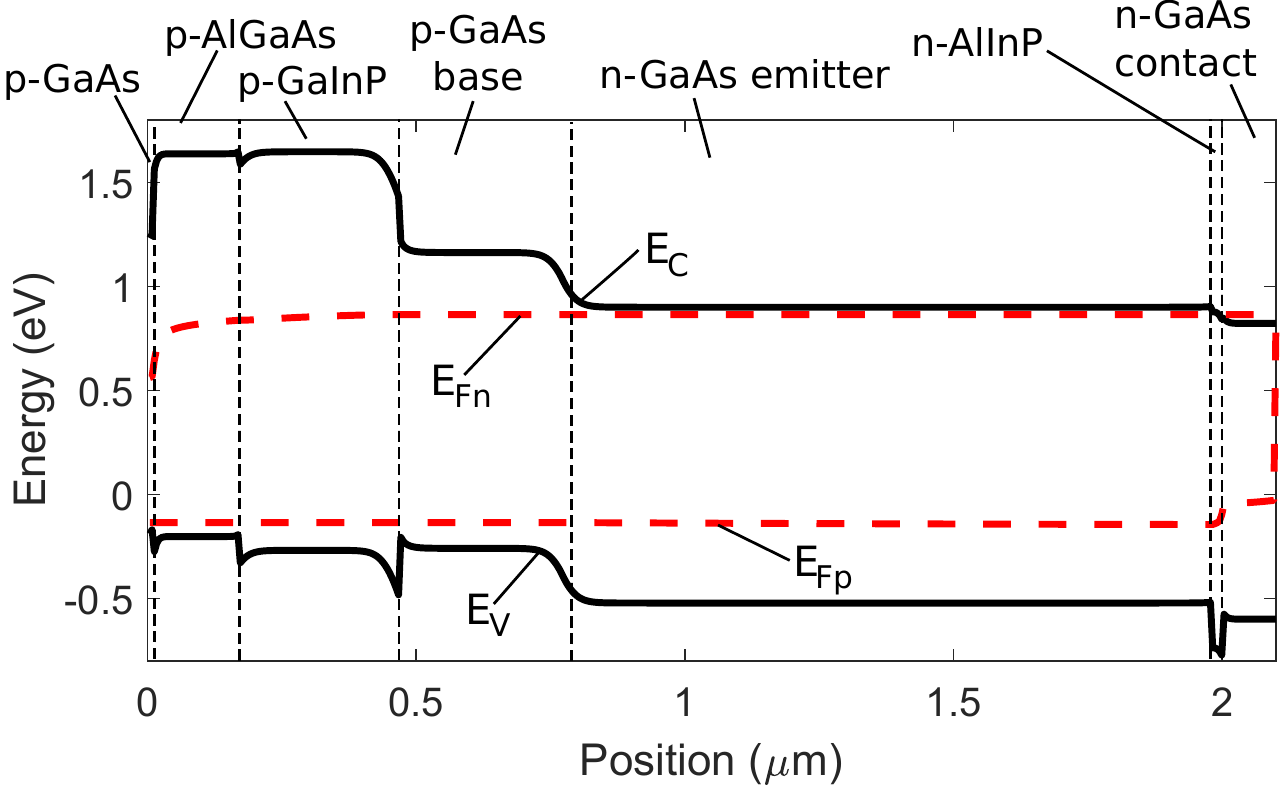}
\caption{Band diagram of the structure simulated with the IRTDD model under illumination and an applied bias of 1.0 V.}
\label{fig:bd_nrel}
\end{figure}

To study the device operation under illumination, Fig.~\ref{fig:bd_nrel} shows the band diagram of the structure under illumination at an applied bias of 1.0 V. The voltage 1.0 V corresponds to the maximum power point (MPP), which is naturally the most interesting operating point of any solar cell. The different layers are also marked in the figure for easier interpretation. The band diagram starts at the p-doped GaAs bottom layer and ends at the n-doped GaAs top layer. The left side of the band diagram is clearly p-type with large potential barriers for electrons to prevent them from leaking to the p-contact. The GaAs pn junction formed by the GaAs base and emitter layers is clearly visible in the middle of Fig.~\ref{fig:bd_nrel}, and the AlInP layer on the right creates a potential barrier to the valence band to prevent holes from leaking to the n-contact. The figure also shows the quasi-Fermi levels $E_{Fn}$ and $E_{Fp}$, whose difference is used as the photon chemical potential in Eq. (\ref{eq:irte}), following Ref. \cite{wurfel_1982}.

\begin{figure}
\centering
\includegraphics[width=\columnwidth]{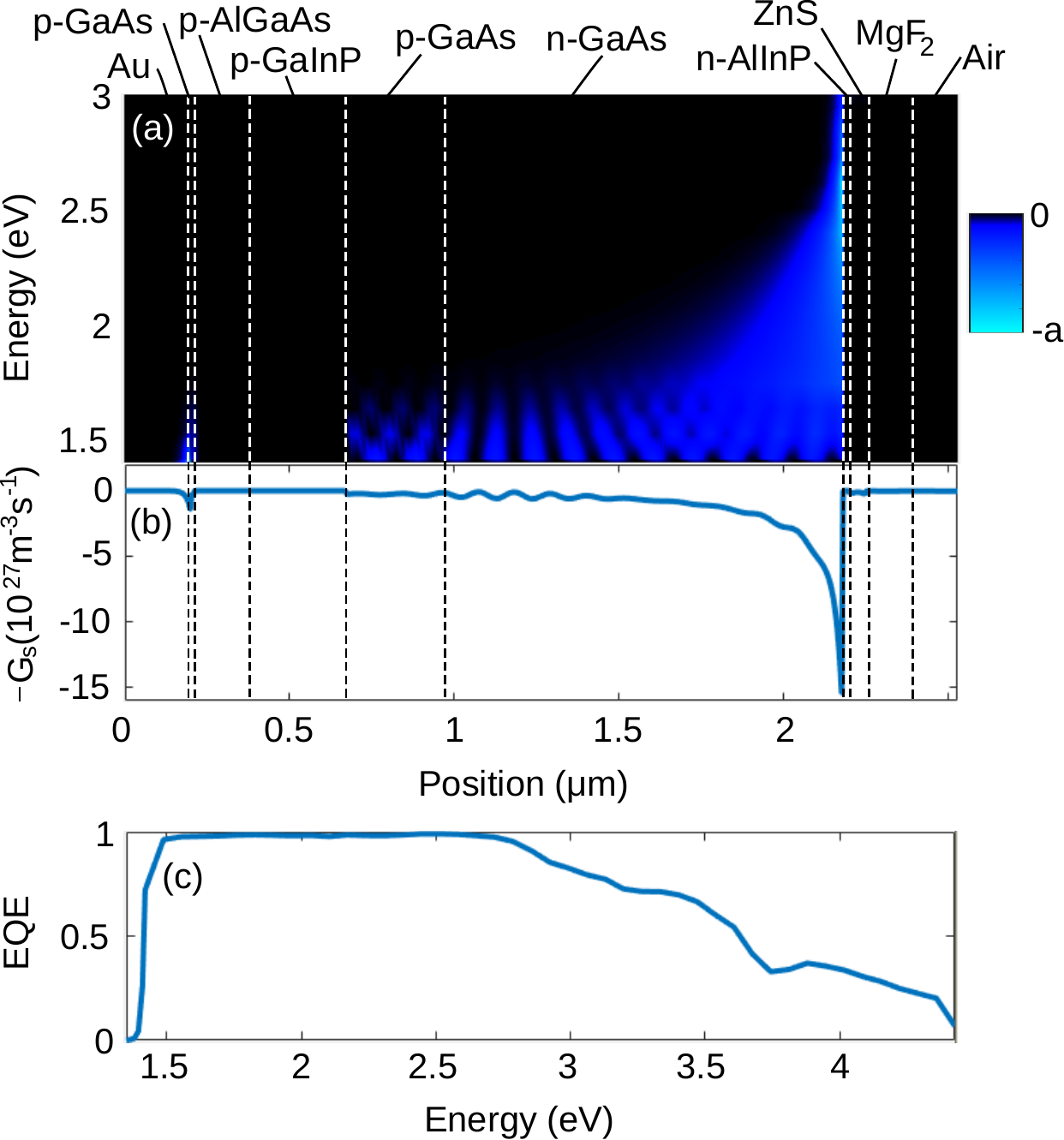}
\caption{(a) Generation rate due to incoming solar photons ($-G_s$) as a function of photon energy and position, (b) generation rate due to incoming solar photons ($-G_s$) as integrated over photon angular frequency, and (c) the estimated external quantum efficiency (EQE) as a function of photon energy, which peaks at almost 100\% due to the AR coating. Here, the ZnS and MgF$_2$ layers constitute the antireflective (AR) coating marked in Fig.~\ref{fig:structure}. In (a), the colorbar value is $a=9.40\times10^{12}$ m$^{-3}$.}
\label{fig:rec_sol}
\end{figure}

For a detailed look at the absorption of sunlight in the structure, Fig.~\ref{fig:rec_sol}(a) shows a colormap of the generation rate due to solar photons as a function of photon energy and position (negative values denoting generation for consistency with the rest of the figures). At low photon energies below roughly 2 eV, generation takes place throughout the GaAs layers due to the modest absorption coefficient and thereby a large contribution also from back-reflected photons. Quite interestingly, there is a clear interference pattern at the lowest photon energies due to the interplay between leftward-propagating photons and the rightward-propagating ones reflected from the gold contact. At photon energies above roughly 2 eV, the generation takes place clearly more towards the top contact on the right side of the figure, as the absorption coefficient of GaAs is large at these photon energies and photons do not reach the lower layers. These trends are also reflected in the solar generation rate integrated over the angular frequency in Fig.~\ref{fig:rec_sol}(b). Furthermore, the external quantum efficiency (EQE) of the solar cell is shown in Fig.~\ref{fig:rec_sol}(c). By comparing the short-circuit current and total carrier generation rate in the solar cell, we have found that the internal quantum efficiency (IQE) is practically 1 for all photon energies in the simulation. Therefore the EQE in Fig.~\ref{fig:rec_sol}(c) is estimated simply by dividing the number of electron-hole pairs created in the GaAs base-emitter layer by the number of photons incident on the solar cell at each wavelength. The EQE is very similar to that reported in the corresponding reference \citenum{steiner_2013}, with the drop at low energies here reflecting the band gap and minor bottom contact absorption, and the decrease at high energies due to reflection losses and minor absorption in the AR coating.

\begin{figure}
\centering
\includegraphics[width=0.9\columnwidth]{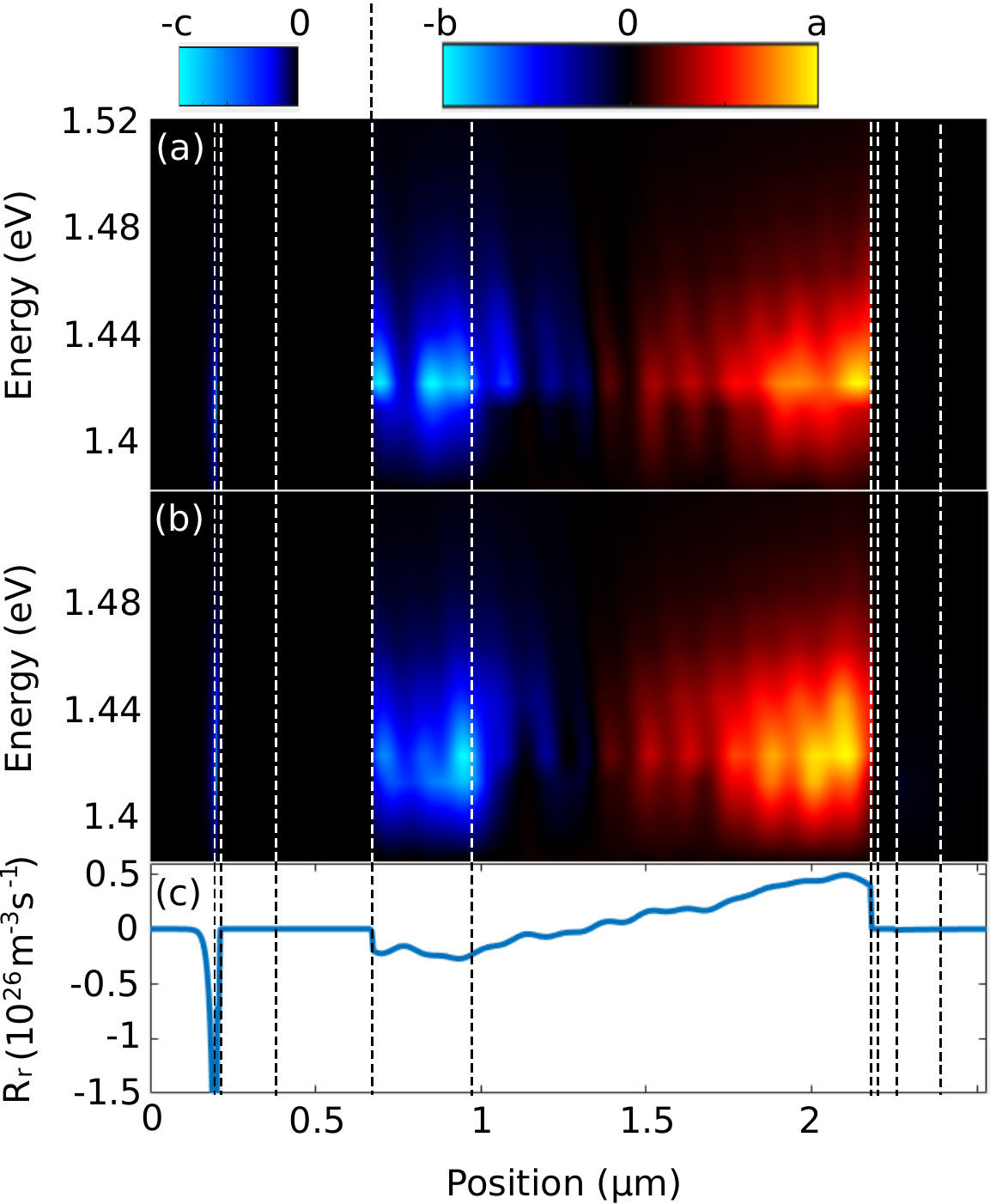}
\caption{Total recombination-generation rate due to internal emission as a function of photon energy and position for (a) TE polarization and (b) TM polarization at an applied bias of 1.0 V under illumination. In (c), both (a) and (b) are integrated over photon angular frequency to give the total RG rate. For this device configuration and bias, there is a very clear spatial separation between regions of net generation and net recombination. Stronger absorption of sunlight towards the top surface creates an imbalance in carrier densities, which is balanced here by recombination at around 2 $\mu$m and a corresponding generation at around 1 $\mu$m. In (a), the colorbar values are $a=3.54\times10^{11}$ m$^{-3}$, $b=2.02\times10^{11}$ m$^{-3}$ and $c=2.53\times10^{12}$ m$^{-3}$, and in (b) they are $a=2.71\times10^{11}$ m$^{-3}$, $b=1.69\times10^{11}$ m$^{-3}$ and $c=2.14\times10^{12}$ m$^{-3}$. By integrating the values in (c) and multiplying by $e$, the total net emission rate is 0.38 mA/cm$^2$, the net rate of photon reabsorption in the GaAs active layers is 0.16 mA/cm$^2$, the rate of photon absorption in the bottom contact layers is 0.10 mA/cm$^2$, and the photon extraction rate is 0.12 mA/cm$^2$.}
\label{fig:rec_E}
\end{figure}

Turning our attention more towards photon recycling, all the figures from now on describe the net recombination-generation (RG) rate $R_r$, representing only those photons that are emitted inside the structure. This will also enable a more direct comparison with the corresponding figures under dark current, which are studied later in Subsection \ref{sec:dark}. To look at the resulting internal RG characteristics, Fig.~\ref{fig:rec_E} shows the RG rate as a function of position and photon energy at an applied bias of 1.0 V under solar illumination. More specifically, Fig.~\ref{fig:rec_E}(a) shows the RG rate for TE polarization, Fig.~\ref{fig:rec_E}(b) shows the same rate for the TM polarization, and Fig.~\ref{fig:rec_E}(c) shows the total rate as a sum of (a) and (b) integrated over photon angular frequency.

Comparing TE and TM in Figs.~\ref{fig:rec_E}(a) and (b), they share the same qualitative characteristics, in which the internal net RG rate is positive (corresponding to net recombination) close to the right side of the figure and negative (corresponding to net generation) towards the left side of the figure. This remarkable photon recycling process is caused by the spatially uneven absorption of solar photons in Fig.~\ref{fig:rec_sol}, which causes $\Delta E_F$ to be roughly a third of the thermal energy $k_BT$ larger next to the top contact than at the p-GaAs base at this operating point, in spite of diffusion that seeks to level off this imbalance. Correspondingly, the maximum of the electron-hole density product $np$ is $5.75\times10^{41}$ m$^{-6}$ next to the top contact, while the minimum is $4.09\times10^{41}$ m$^{-6}$ at the p-GaAs base. Were radiative recombination calculated instead using the conventional model with a radiative recombination coefficient, this would correspond to a radiative rate being roughly 40 \% larger close to the top contact than at the p-GaAs base. Nevertheless, this internal optical energy transfer process is a direct consequence of the spatially uneven electrical excitation in the structure. As such, it would be impossible to reveal quantitatively without access to a direct and fully self-consistent electro-optical modeling framework. The internal recombination-generation in Fig.~\ref{fig:rec_E} mostly takes place at photon energies notably below 1.5 eV, as the Fermi-Dirac distributions decay fast as a function of the recombination energy. There are also visible quantitative differences between the TE and TM RG rates in Fig.~\ref{fig:rec_E} due to wave-optical effects. For example, the maxima and minima of the RG rate occur at somewhat different locations in (a) and (b) due to the different mode structure of the TE and TM polarizations. The caption of Fig.~\ref{fig:rec_E} also lists the total internal net recombination rate and the different generation rates multiplied by $e$. Based on those, the internal RG rates are still very small as compared to the total current at this MPP, as one would expect. This strongly suggests that in spite of the new internal optical energy transfer process revealed by the IRTDD, the device-level characteristics of the solar cell can still be quite reliably modeled with typical analytical approximations, e.g., by using a properly calibrated radiative recombination coefficient \cite{niemeyer_2019}.

\begin{figure}
\centering
\includegraphics[width=0.85\columnwidth]{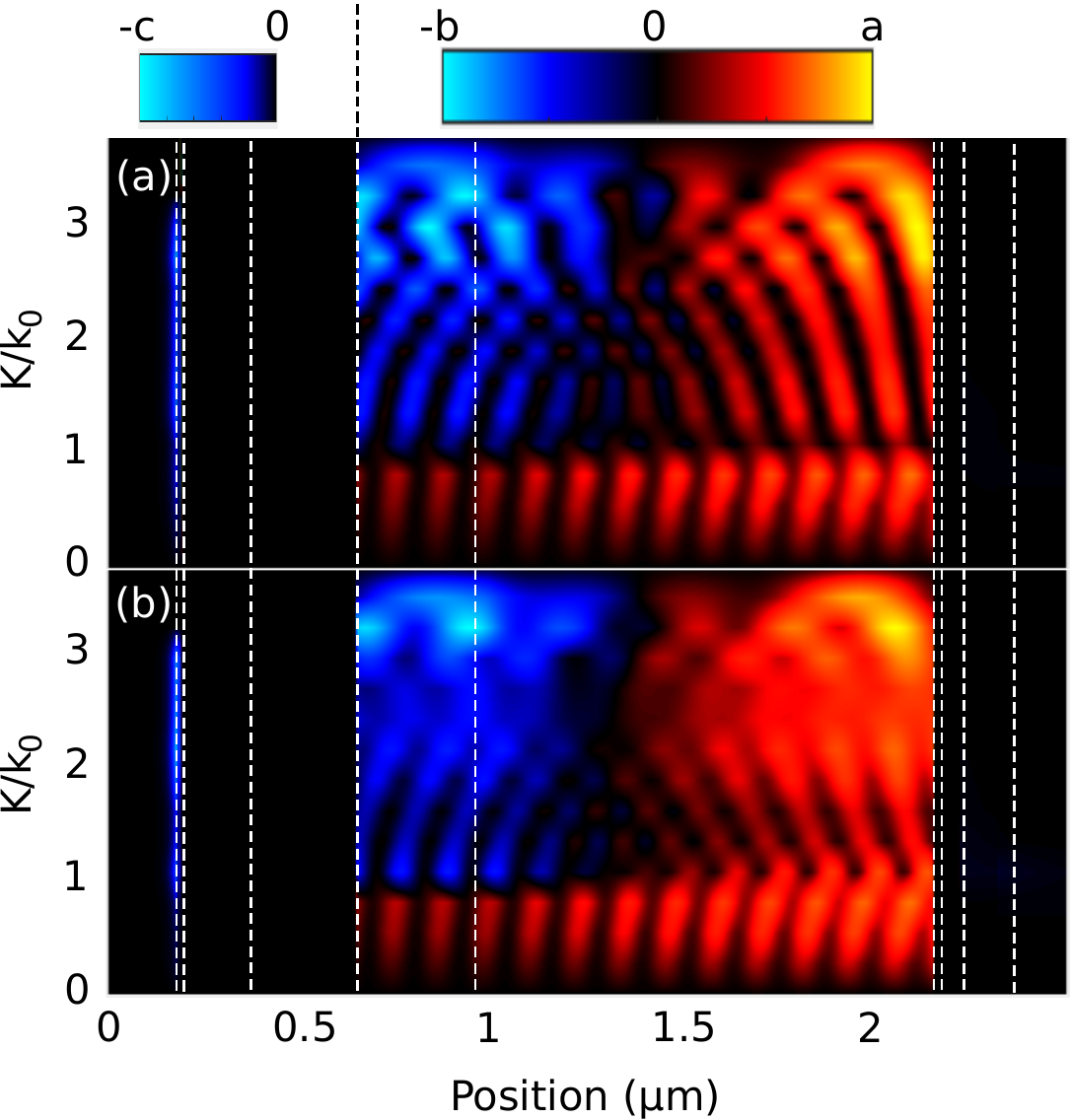}
\caption{Recombination-generation rate due to internal emission as a function of $K$ and position at photon energy 1.43 eV at an applied bias of 1.0 V under illumination for (a) TE and (b) TM polarization. The rates include the integration factor $2\pi K$. In (a), the colorbar values are $a=0.0048$ m$^{-2}$, $b=0.0036$ m$^{-2}$ and $c=0.0183$ m$^{-2}$, and in (b) they are 0.0039 m$^{-2}$, $b=0.0032$ m$^{-2}$ and $c=0.0243$ m$^{-2}$.}
\label{fig:rec_E1p43}
\end{figure}

To break down the optical energy transfer processes in more detail at one relevant photon energy, Fig.~\ref{fig:rec_E1p43} shows the RG rate as a function of $K$ at 1.43 eV (a value close to the GaAs bandgap energy) for (a) TE and (b) TM polarization. To compare with the previous figure, the row at 1.43 eV in Fig.~\ref{fig:rec_E} represents the integral of Fig.~\ref{fig:rec_E1p43} over all $K$. Above $K/k_0=1$, Fig.~\ref{fig:rec_E1p43} naturally exhibits the same trend in the spatial distribution as Fig.~\ref{fig:rec_E}, where the RG rate is positive towards the top contact on the right side and negative towards the bottom contact on the left side. Below $K/k_0=1$, the RG rate is positive throughout the active GaAs layers, corresponding mostly to photons extracted through the top contact. Moreover, here the interference effects are clearly visible as spatial oscillations in the RG rate due to constructive and destructive interferences and possibly also due to the Purcell effect, demonstrating the effects that become crucial when optimizing the electro-optical transport e.g. within resonant cavities or through near fields. The RG rates approach zero roughly at $K/k_0=3.5$, when no more propagating modes exist in the structure. Again, it is noteworthy that these rather complex internal optical energy transfer processes are practically impossible to calculate without a self-consistent model of emission and photon recycling applicable for unevenly excited structures. As such, they have not previously been reported due to the difficulty to simultaneously account for all the underlying physical processes. We expect such spectrally and directionally resolved understanding of photon recycling processes to be especially beneficial for understanding the internal electrical and optical properties of various thin-film devices, such as the emerging ultra-thin GaAs photovoltaic devices \cite{chen_2019,yang_2014,gai_2017,vaneerden_2020}, and to complement existing advanced modeling and characterization frameworks of multi-junction solar cells \cite{timo_2020,xu_2019}. 

\begin{figure}
\centering
\includegraphics[width=0.85\columnwidth]{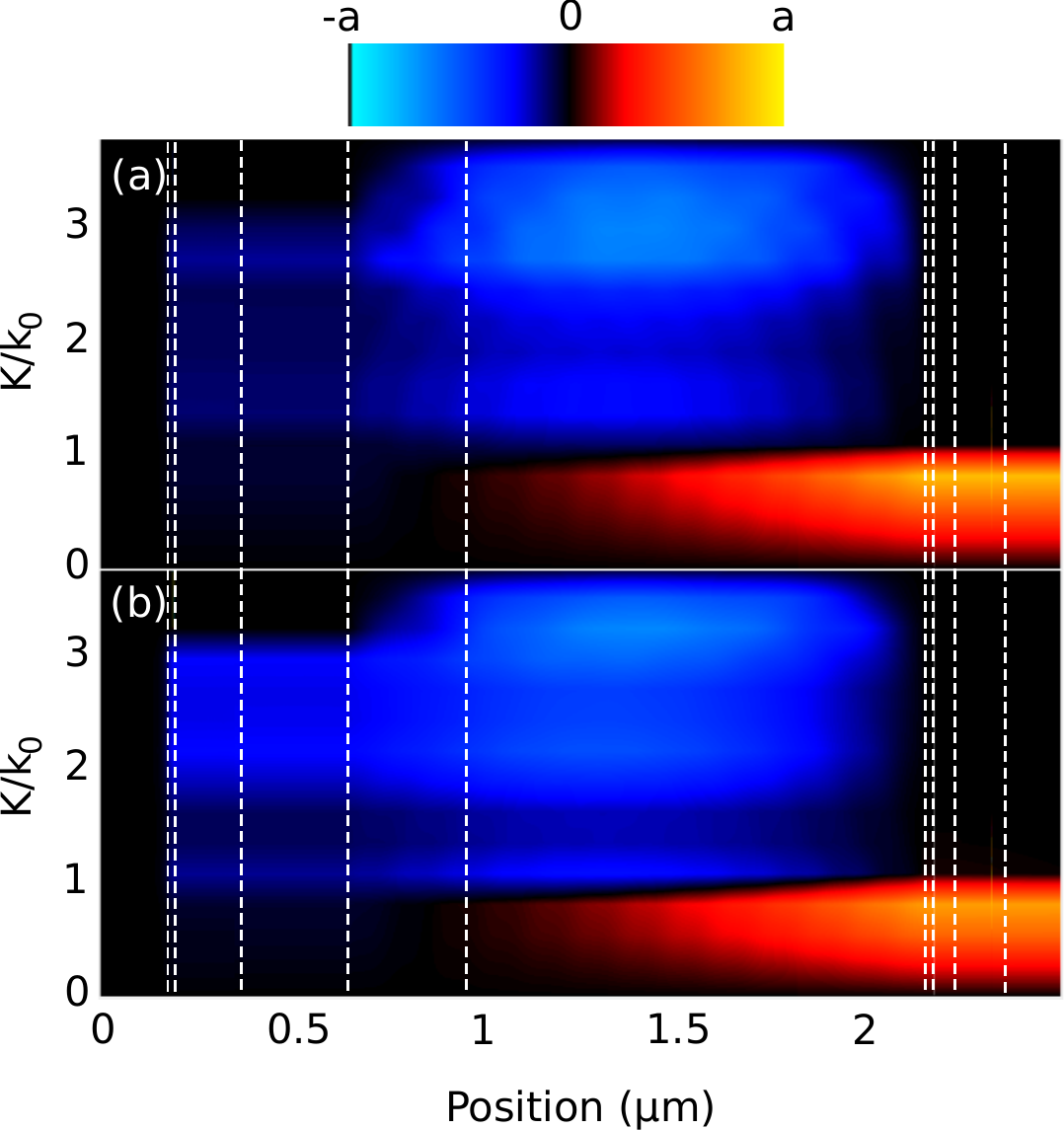}
\caption{Spectral radiance $S\times2\pi K$ as a function of $K$ and position at photon energy 1.43 eV at an applied bias of 1.0 V under illumination for (a) TE and (b) TM polarization. In both (a) and (b), the colorbar value is $a=4\times10^{-28}$ J/m.}
\label{fig:poynt_E1p43}
\end{figure}

To study how the optical power propagates within the solar cell structure studied here, Fig.~\ref{fig:poynt_E1p43} shows the $K$-resolved spectral radiance of internally emitted photons as a function of position and $K$ at the same photon energy 1.43 eV for (a) TE and (b) TM polarization, again for 1.0 V. It can be seen that in the escape cone determined by $K/k_0\leq1$, there is a clear positive optical power flow out of the device in the bottom right corner of the figures. However, based on a closer inspection of the results, it is only around 0.4 \% of the absorbed solar optical power, further indicating that the structure is still operating fully as a solar cell at this bias. At larger $K$ values, the power escaping the structure becomes zero, but the spectral radiance is negative in the middle of the GaAs layers. This corresponds to the optical energy transfer process studied in the previous figures, where light is emitted close to the top contact on the right side and absorbed in the lower GaAs layers and in the bottom contact layers on the left side due to the imbalance in $\Delta E_F$ created by sunlight. This illustrates once more how the interplay of photons and charge carriers captured by the self-consistent simulation evens up part of the imbalance in the carrier densities created by solar absorption.

\subsection{Characteristics under dark current} \label{sec:dark}

\begin{figure}
\centering
\includegraphics[width=0.9\columnwidth]{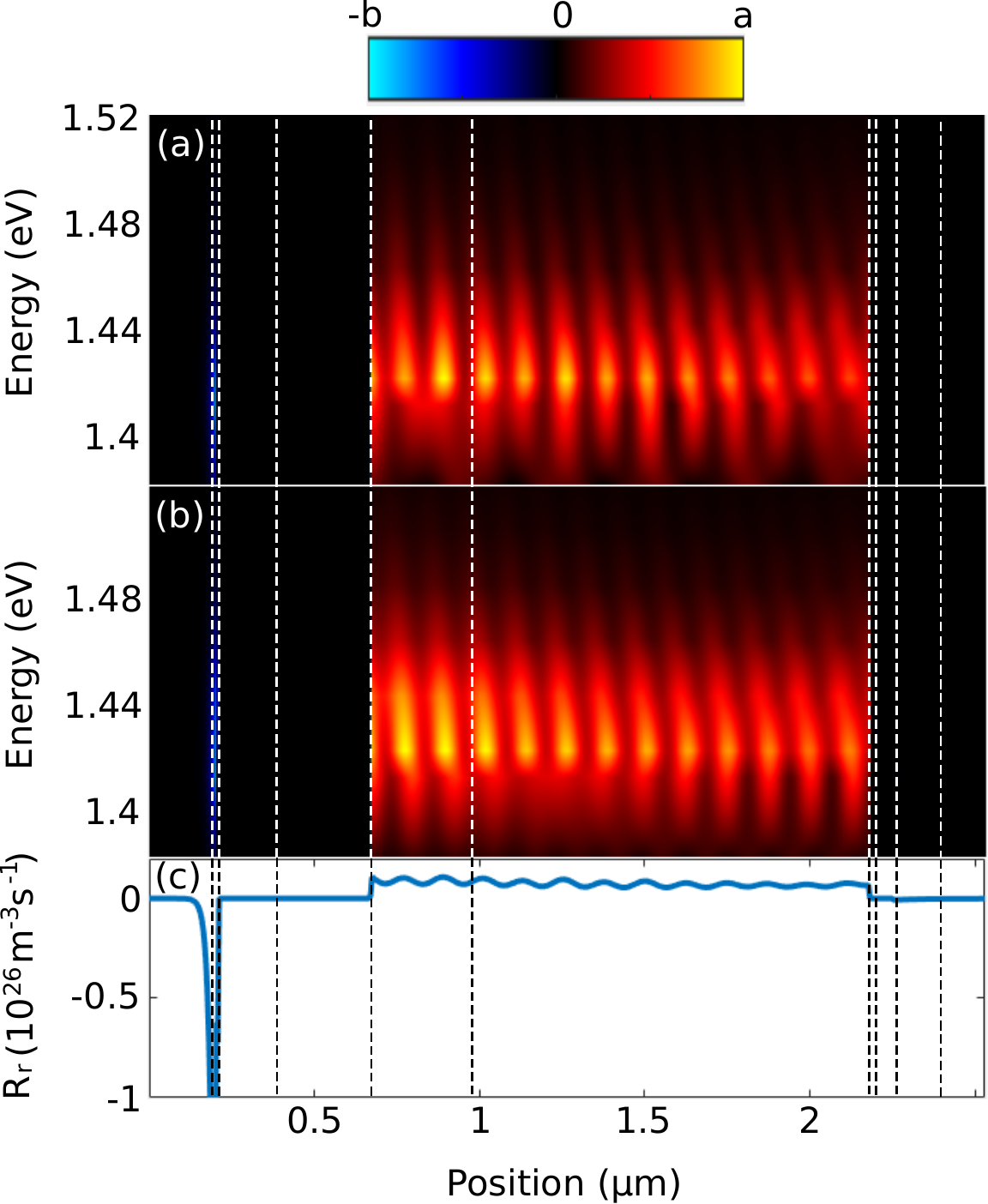}
\caption{Information in Fig.~\ref{fig:rec_E} repeated for dark current conditions: recombination-generation rate due to internal emission as a function of photon energy and position at an applied bias of 1.0 V under dark current conditions, including all propagation directions: (a) TE and (b) TM polarization. In (c), both (a) and (b) are integrated over photon angular frequency to give the total RG rate. In (a), the colorbar values are $a=7.08\times10^{10}$ m$^{-3}$ and $b=2.13\times10^{12}$ m$^{-3}$ and in (b), they are $a=6.64\times10^{10}$ m$^{-3}$ and $b=1.80\times10^{12}$ m$^{-3}$. By integrating the values in (c) and multiplying by $e$, the total net emission rate is 0.18 mA/cm$^2$, the net rate of photon reabsorption in the GaAs active layers is 0, the rate of photon absorption by the bottom contact layers is 0.08 mA/cm$^2$, and the photon extraction rate is 0.10 mA/cm$^2$.}
\label{fig:rec_E_dark}
\end{figure}

In this Section, we repeat the internal RG results for the dark current case to shortly illustrate the physical insight provided by the IRTDD model at another commonly used operation mode in solar cell research. The information of Fig.~\ref{fig:rec_E} is repeated for dark current conditions in Fig.~\ref{fig:rec_E_dark}, which again shows the RG rate density as a function of photon energy and position for (a) TE and (b) TM at the bias voltage of 1.0 V. Again, Fig.~\ref{fig:rec_E_dark}(c) shows the total rate as a sum of (a) and (b) integrated over photon angular frequency. Interestingly, the result is very different from the one shown in Fig.~\ref{fig:rec_E} calculated at a similar voltage under solar illumination. In the GaAs region of Fig.~\ref{fig:rec_E_dark}, we only see positive net RG rates, whereas in Fig.~\ref{fig:rec_E} under illumination, there was also a negative RG rate towards the bottom of the GaAs layer on the left, corresponding to net generation there. In the dark current case shown in Fig.~\ref{fig:rec_E_dark}, there is no clear imbalance in the electron-hole densities created by solar absorption. Therefore net emission is created through the process of radiation out from the structure and through photons that are absorbed by the bottom contact. As with Fig.~\ref{fig:rec_E}, the caption also lists the total internal net recombination rate and the generation rates multiplied by $e$. Comparing the captions of Fig. \ref{fig:rec_E_dark} (dark current) and Fig.~\ref{fig:rec_E} (under illumination), one can see that there indeed is no net generation in GaAs under dark current and that the rate of bottom contact absorption and photon extraction are only slightly smaller than in the illuminated case. On the other hand, the internal net recombination rate is notably larger under illumination than under dark current at the MPP, as one would expect for a high-efficiency solar cell.

\begin{figure}
\centering
\includegraphics[width=0.85\columnwidth]{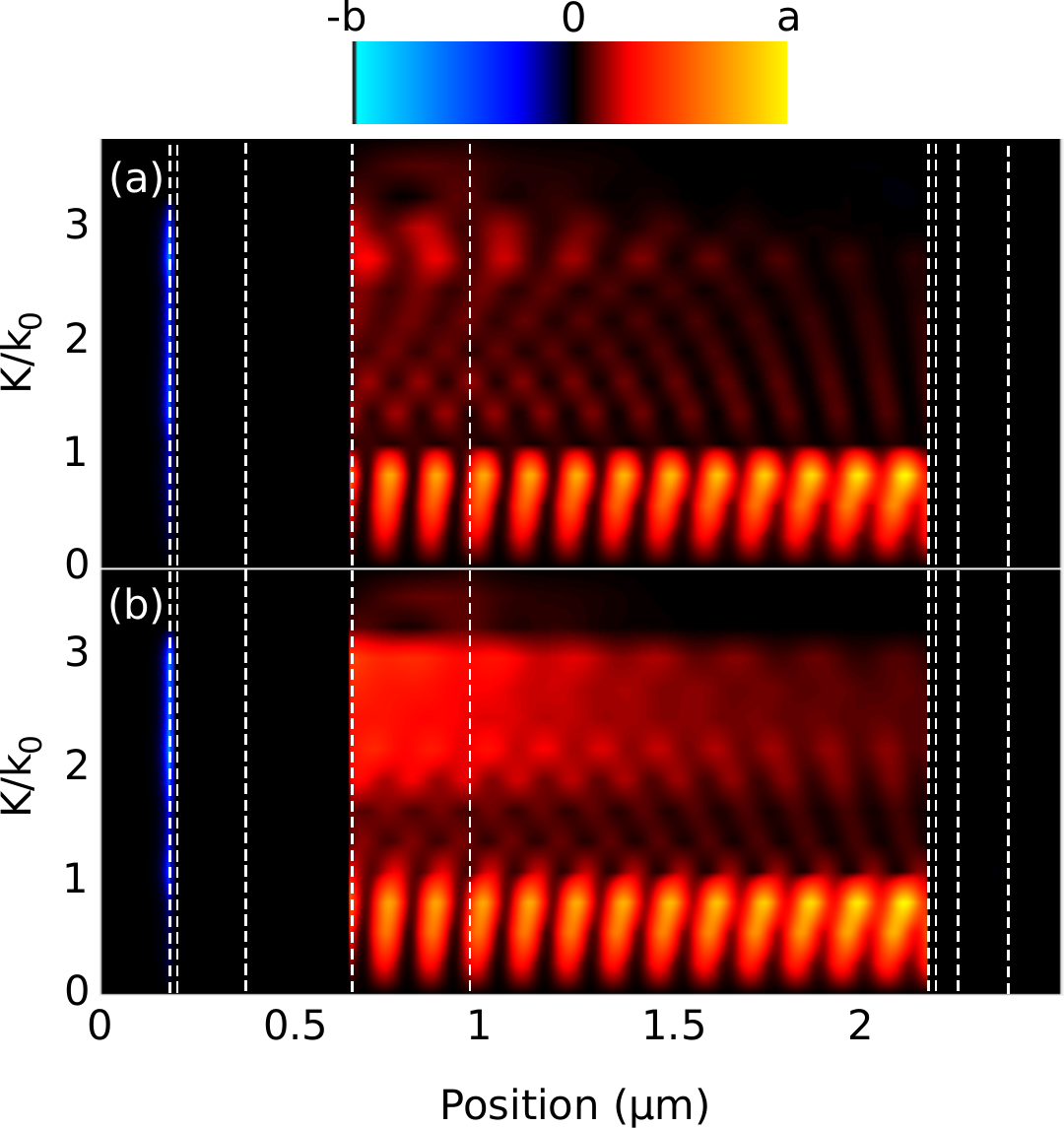}
\caption{Information in Fig.~\ref{fig:rec_E1p43} repeated for dark current conditions: recombination-generation rate due to internal emission as a function of $K$ and position at photon energy 1.43 eV at an applied bias of 1.0 V under dark current conditions for (a) TE and (b) TM polarization. The rates include the integration factor $2\pi K$. In (a), the colorbar values are $a=0.0017$ m$^{-2}$ and $b=0.0161$ m$^{-2}$ and in (b), they are $a=0.0015$ m$^{-2}$ and $b=0.0212$ m$^{-2}$.}
\label{fig:rec_E1p43_dark}
\end{figure}

\begin{figure}
\centering
\includegraphics[width=0.85\columnwidth]{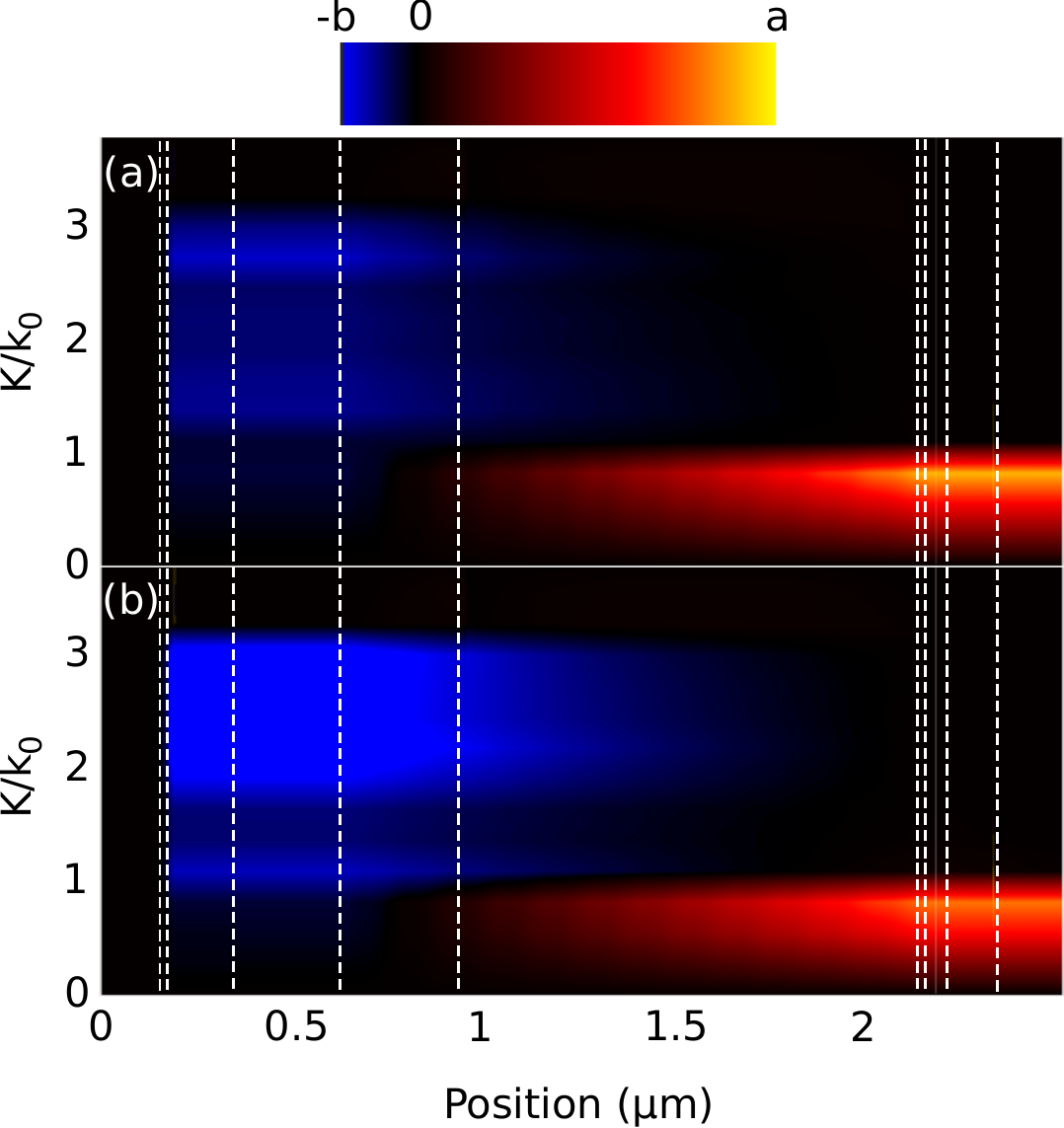}
\caption{Information in Fig.~\ref{fig:poynt_E1p43} repeated for dark current conditions: spectral radiance $S\times2\pi K$ as a function of $K$ and position at photon energy 1.43 eV at an applied bias of 1.0 V under dark current conditions for (a) TE and (b) TM polarization. In both (a) and (b), the colorbar values are $a=3\times10^{-28}$ J/m and $b=0.65\times10^{-28}$ J/m.}
\label{fig:poynt_E1p43_dark}
\end{figure}

The RG rate in dark current conditions can be studied in more detail with help of Fig.~\ref{fig:rec_E1p43_dark}, which repeats the RG rates of Fig.~\ref{fig:rec_E1p43} in dark as a function of $K$ and position for (a) TE and (b) TM for a single energy 1.43 eV, again at 1.0 V. It can be seen that the $K$-resolved RG rates are largest in the modes up to $K/k_0=1$, corresponding primarily to photons that radiate out from the structure and partly to photons absorbed by the bottom contact. Above $K/k_0=1$, the slightly smaller RG rates correspond to photons that are absorbed by the bottom contact on the left. This observation is complemented by Fig.~\ref{fig:poynt_E1p43_dark}, which repeats Fig.~\ref{fig:poynt_E1p43} in dark for (a) TE and (b) TM modes as a function of $K$ and position. Here the spectral radiance has its clearly largest $K$-resolved values at $K/k_0<1$, corresponding to photons exiting through the AR coating on the right side of the figure and at angles close to escape cone boundary in particular.

\subsection{Final remarks on the results}

The analysis in the previous Subsections demonstrated how the IRTDD model can be used to accurately calculate the internal recombination-generation characteristics for essentially any planar device structure. The IRTDD model was introduced by applying it specifically to a well-known and widely studied device, both to confirm that it gives the expected device-level results and to demonstrate the added physical insight that can be gained. In particular, the optical energy transfer effects illustrated in Figs. \ref{fig:rec_E}-\ref{fig:poynt_E1p43} represent new microscopic insight into one of today's most important renewable energy technologies that can only be obtained in this level of detail with such accurate self-consistent electro-optical models applicable for unevenly excited structures. While these effects do not crucially alter the macroscopic performance of the device studied here, they are expected to be pivotal in selected emerging devices where the optical energy transport primarily takes place e.g. within cavities or through near fields \cite{sadi_2020,fan_2020,zhu_2019}. On the other hand, performing the IRTDD calculations on a typical solar cell here provided the important confirmation that despite the peculiar optical energy transfer processes revealed, the device-level characteristics are still primarily determined by effects included in analytical approximations typically used in full-device studies.

Through its rigorous solution of Maxwell's equations and the use of fluctuational electrodynamics, the IRTDD model represents a direct way to calculate light emission in arbitrary planar structures \cite{wang_2011}. This provides an elegant and accurate solution of the internal emission and absorption processes even in the case of a spatially varying quasi-Fermi level separation. It can be insightful to compare the IRTDD model conceptually with indirect methods based on detailed balance considerations of absorption and emission, which are frequently used for studying similar topics \cite{ren_2017}. If the quasi-Fermi levels are constant throughout the emitting materials and no parasitic optical losses are present in the emitter cavity, emission and absorption can be related using Kirchhoff's law to calculate the absorption and emission and reconstruct the interference effects. However, in the presence of nonuniform excitation or parasitic contact losses, it generally becomes necessary to use direct emission models based on fluctuational electrodynamics rather than resorting to Kirchhoff's law and absorptivities \cite{wang_2011}. For the solar cell studied here, the indirect method likely produces a good picture of the total emission. However, due to the possible shortcomings of Kirchhoff's law in the case of spatially varying quasi-Fermi levels and energy losses associated with guided modes, it is important to have tools also for direct simulation.

\section{Conclusions}

To improve the theoretical insight into various thin-film optoelectronic devices, we introduced an accurate and fully self-consistent model of photon and charge carrier transport applicable to planar devices. The model was based on connecting the drift-diffusion formalism of carrier transport and fluctuational electrodynamics of photon transport by making use of the newly developed interference-extended radiative transfer equations. To demonstrate the thus obtained IRTDD modeling tool, we applied it to study the device characteristics and internal optical properties of a thin-film GaAs solar cell. The results indicate that the IRTDD model not only reproduces the expected device characteristics, but also directly provides detailed physical insight into complex nonlocal effects such as photon recycling, which is a relevant process particularly in high-efficiency thin-film solar cells. The IRTDD framework presented here is general, and it could be implemented in a wide range of planar resonant devices for detailed studies of photon recycling, luminescent coupling and other complex forms of electro-optical interaction.

\section*{Acknowledgments}

We acknowledge financial support from the Academy of Finland (grant number 315403), the European Research Council's Horizon 2020 research and innovation programme (grant agreement No 638173), and the Horizon 2020 Marie Sk{\l}odowska-Curie Actions individual fellowship under contract No 846218. Computational resources provided by the Aalto Science-IT project were used to carry out the simulations of this paper.

\appendix

\section{Modified photon numbers and related quantities} \label{sec:appendixa}

To use the IRT model in solar cells and other devices, it is convenient if the incoming photon numbers can be set to zero at the outer boundaries of the computational domain as a boundary condition. To this end, we slightly modify the definition of the interference density of states (IFDOS) as compared to the one used in Ref. \citenum{partanen_2017_sr}. A more complete description of the full formalism can be found in our previous works, here most importantly Refs. \citenum{partanen_2017_sr,partanen_2017,kivisaari_2018}. Using the notation of Ref. \citenum{partanen_2017_sr}, we begin by writing the spectral radiance as
\begin{widetext}
\begin{equation}
\langle\hat{S}_{\sigma}(z,K,\omega)\rangle=\hbar\omega v_{\sigma}(z,K,\omega)\int_{-\infty}^{\infty}\rho_{IF,\sigma}(z,K,\omega,z')\langle\hat{\eta}(z',\omega)\rangle dz',
\end{equation}
where the generalized velocities of light are given by
\begin{equation}
v_{TE}=\frac{2ck_0|\mu|\Re(k_z/\mu)}{|k|^2+|k_z|^2+K^2},\hspace{2cm}v_{TM}=\frac{2ck_0|\varepsilon|\Re(k_z/\varepsilon)}{|k|^2+|k_z|^2+K^2}.
\end{equation}
These changes in the velocities are balanced by scaling the IFDOSs with a factor $\zeta_{v,\sigma}=c/(v_\sigma n_r)$. The value of the spectral radiance is unchanged, but the IFDOSs and the related propagating photon numbers are slightly modified as specified below. Making use of the photon numbers propagating in the positive or negative direction instead, the spectral radiance is written as in Eq.~(\ref{eq:irte}). With this choice of $v_{\sigma}$, we obtain that outside the simulation domain, $\alpha_+=\mbox{const.}$ in the negative direction, $\alpha_-=\mbox{const.}$ in the positive direction, $\beta_+=0$ in the negative direction, and $\beta_-=0$ in the positive direction. In addition, the mentioned constant values of $\alpha_+$ and $\alpha_-$ are zero in lossless media. This allows one to set the input photon numbers to zero in all lossy and lossless cases when there are no sources outside. Using the mentioned scaling of IFDOSs of Ref.~\citenum{partanen_2017_sr} with a factor $\zeta_{v,\sigma}=c/(v_\sigma n_r)$, the definitions of the IFDOSs are here given by
\begin{equation}
\rho_{IF,TE}(z,K,\omega,z')=-\frac{\omega^3}{2\pi^3c^3v_{TE}}\Im\left[\varepsilon_i'g_{ee}^{11}g_{me}^{21*}+\mu_i'g_{mm}^{22}g_{em}^{12*}+\mu_i'g_{mm}^{23}g_{em}^{13*}\right],
\end{equation}
\begin{equation}
\rho_{IF,TM}(z,K,\omega,z')=\frac{\omega^3}{2\pi^3c^3v_{TM}}\Im\left[\mu_i'g_{mm}^{11}g_{em}^{21*}+\varepsilon_i'g_{ee}^{22}g_{me}^{12*}+\varepsilon_i'g_{ee}^{23}g_{me}^{13*}\right].
\end{equation}
These IFDOSs and their position derivatives together with the other density of state terms, given in the Supplemental Material of Ref. \citenum{partanen_2017_sr}, allow simplifying the damping and scattering coefficients in the general equations of Ref. \citenum{partanen_2017_sr} to their final forms given by
\begin{equation}
\alpha_{\pm,TE}=\frac{k_rk_0}{\rho_{TE}/\rho_0}\Im\left[\left(\frac{\zeta_{e,TE}}{\zeta_{v,TE}}+\zeta_{v,TE}\varepsilon_i\right)g_{ee}^{11}+\left(\frac{\zeta_{m,TE}}{\zeta_{v,TE}}+\zeta_{v,TE}\mu_i\right)g_{mm}^{22}\pm\zeta_{ex,TE}(g_{me}^{21}-g_{em}^{12})+\zeta_{v,TE}\mu_i\frac{\mu^2}{|\mu|^2}g_{mm}^{33}\right],
\end{equation}
\begin{equation}
\beta_{\pm,TE}=\frac{k_rk_0}{\rho_{TE}/\rho_0}\Im\left[\left(\frac{\zeta_{e,TE}}{\zeta_{v,TE}}-\zeta_{v,TE}\varepsilon_i\right)g_{ee}^{11}+\left(\frac{\zeta_{m,TE}}{\zeta_{v,TE}}-\zeta_{v,TE}\mu_i\right)g_{mm}^{22}\pm\zeta_{ex,TE}^*(g_{me}^{21}-g_{em}^{12})-\zeta_{v,TE}\mu_i\frac{\mu^2}{|\mu|^2}g_{mm}^{33}\right],
\end{equation}
\begin{equation}
\alpha_{\pm,TM}=\frac{k_rk_0}{\rho_{TM}/\rho_0}\Im\left[\left(\frac{\zeta_{m,TM}}{\zeta_{v,TM}}+\zeta_{v,TM}\mu_i\right)g_{mm}^{11}+\left(\frac{\zeta_{e,TM}}{\zeta_{v,TM}}+\zeta_{v,TM}\varepsilon_i\right)g_{ee}^{22}\pm\zeta_{ex,TM}(g_{em}^{21}-g_{me}^{12})+\zeta_{v,TM}\varepsilon_i\frac{\varepsilon^2}{|\varepsilon|^2}g_{ee}^{33}\right],
\end{equation}
\begin{equation}
\beta_{\pm,TM}=\frac{k_rk_0}{\rho_{TM}/\rho_0}\Im\left[\left(\frac{\zeta_{m,TM}}{\zeta_{v,TM}}-\zeta_{v,TM}\mu_i\right)g_{mm}^{11}+\left(\frac{\zeta_{e,TM}}{\zeta_{v,TM}}-\zeta_{v,TM}\varepsilon_i\right)g_{ee}^{22}\pm\zeta_{ex,TM}^*(g_{em}^{21}-g_{me}^{12})-\zeta_{v,TM}\varepsilon_i\frac{\varepsilon^2}{|\varepsilon|^2}g_{ee}^{33}\right].
\end{equation}
Here, the parameters and their equations are given in the Supplemental Material of Ref. \citenum{partanen_2017_sr}.

\begin{table}[b]
\centering
\caption{Parameters used in the DD simulations of electron-hole transport \cite{vurgaftman_2001,levinshtein_1996}. The band offsets $dE_c$ and $dE_v$ are given with respect to the band edges in GaAs.}
\begin{tabular}{c|c|c|c|c|c|c|c|c}
\hline
\hline
Property & GaAs emitter & GaAs base & GaAs p-contact & GaAs n-contact & AlGaAs & InGaP & AlInP & Unit \\
\hline
$N$ & $10^{23}$ & $-10^{23}$ & $-10^{24}$ & $10^{24}$ & $-10^{24}$ & $-10^{23}$ & $10^{23}$ & 1/m$^3$ \\
\hline
$\varepsilon$ & \multicolumn{4}{c|}{12.90} & 12.05 & 11.79 & 12.02 & $\varepsilon_0$ \\
$E_g$ & \multicolumn{4}{c|}{1.43} & 1.84 & 1.91 & 1.62 & eV \\
$dE_v$ & \multicolumn{4}{c|}{0} & -0.16 & -0.31 & -0.22 & eV \\
$dE_c$ & \multicolumn{4}{c|}{0} & 0.26 & 0.19 & -0.03 & eV \\
$\mu_n$ & \multicolumn{4}{c|}{8500} & 2300 & 2800 & 5400 & cm$^2$/(Vs) \\
$\mu_p$ & \multicolumn{4}{c|}{400} & 146 & 175 & 200 & cm$^2$/(Vs) \\
$m_e^*$ & \multicolumn{4}{c|}{0.067} & 0.092 & 0.093 & 0.074 & $m_0$ \\
$m_h^*$ & \multicolumn{4}{c|}{0.61} & 0.66 & 0.77 & 0.84 & $m_0$ \\
\hline
$A$ & \multicolumn{7}{c|}{3$\times10^6$} & 1/s \\
$C$ & \multicolumn{7}{c|}{7$\times10^{-41}$} & m$^6$/s \\
$B$ & \multicolumn{7}{c|}{N/A} & m$^3$/s \\
\hline
\hline
\end{tabular}
\label{tab:params}
\end{table}

\section{Simulation parameters and other simulation details} \label{sec:appendixb}

Parameters used in the DD simulations of electron-hole transport are listed in Table \ref{tab:params}. The radiative recombination coefficient $B$ is not used in the final IRTDD simulations, but an initial value of $7.2\times10^{-16}$ m$^3$/s was used to get initial conditions. The $A$ and $C$ recombination coefficients are set to selected reasonable values for GaAs, and with these values the open-circuit voltage is still determined by radiative recombination. The $A$ and $C$ coefficients are the same for all the materials for simplicity, as recombination takes place predominantly in GaAs due to the relatively low injection levels.

For the optical simulations, we use photon energy-dependent complex dielectric functions reported in the literature for Au \cite{palik_gold}, GaAs \cite{palik_1985}, AlGaAs \cite{palik_algaas}, GaInP \cite{schubert_1995}, AlInP \cite{linnik_2002}, ZnS \cite{ozaki_1993}, and MgF$_2$ \cite{rodriguez_2017}, with help from \cite{polanskiy}. The relative permeability is assumed to be 1 in all materials at all photon energies.

\end{widetext}


\begin{thebibliography}{99}

\bibitem{green_2016}
M. A. Green, Commercial progress and challenges for photovoltaics, Nat. Energy \textbf{1}, 15015 (2016).

\bibitem{cho_2017}
J. Cho, J. H. Park, J. K. Kim, and E. F. Schubert, White light-emitting diodes: History, progress, and future, Laser Photonics Rev. \textbf{11}, 1600147 (2017).

\bibitem{bayvel_2016}
P. Bayvel, R. Maher, T. Xu, G. Liga, N. A. Shevchenko, D. Lavery, A. Alvarado, and R. I. Killey, Maximizing the optical network capacity, Phil. Trans. R. Soc. A. \textbf{374}, 20140440 (2016). 

\bibitem{cappelluti_2018}
F. Cappelluti, D. Kim, M. van Eerden, A. P. C\'edola, T. Aho, G. Bissels, F. Elsehrawy, J. Wu, H. Liu, P. Mulder, G. Bauhuis, J. Schermer, T. Niemi, and M. Guina, Light-trapping enhanced thin-film III-V quantum dot solar cells fabricated by epitaxial lift-off, Sol. Energ. Mat. Sol. C. \textbf{181}, 83 (2018).

\bibitem{sai_2019}
H. Sai, T. Oku, Y. Sato, M. Tanabe, T. Matsui, and K. Matsubara, Potential of very thin and high-efficiency silicon heterojunction solar cells, Prog. Photovolt.: Res. Appl. \textbf{27}, 1061 (2019).

\bibitem{eerden_2019}
M. van Eerden, G. J. Bauhuis, P. Mulder, N. Gruginskie, M. Passoni, L. C. Andreani, E. Vlieg, and J. J. Schermer, A facile light-trapping approach for ultrathin GaAs solar cells using wet chemical etching, Prog. Photovolt.: Res. Appl. \textbf{28}, 200 (2020).

\bibitem{massiot_2014}
I. Massiot, N. Vandamme, N. Bardou, C. Dupuis, A. Lema\^itre, J.-F. Guillemoles, and S. Collin, Metal Nanogrid for Broadband Multiresonant Light-Harvesting in Ultrathin GaAs Layers, ACS Photon. \textbf{1}, 878 (2014).

\bibitem{metaferia_2019}
W. Metaferia, K. L. Schulte, J. Simon, S. Johnston, and A. J. Ptak, Gallium arsenide solar cells grown at rates exceeding 300 $\mu$mh$^{-1}$ by hydride vapor phase epitaxy, Nat. Commun. \textbf{10}, 3361 (2019).

\bibitem{schermer_2005}
J. J. Schermer, P. Mulder, G. J. Bauhuis, P. K. Larsen, G. Oomen, and E. Bongers, Thin-film GaAs epitaxial lift-off solar cells for space applications, Prog. Photovolt.: Res. Appl. \textbf{13}, 587 (2005).

\bibitem{wilkins_2015}
M. Wilkins, C. E. Valdivia, A. M. Gabr, D. Masson, S. Fafard, and K. Hinzer, Luminescent coupling in planar opto-electronic devices, J. Appl. Phys. \textbf{118}, 143102 (2015).

\bibitem{walker_2015}
A. W. Walker, O. H\"ohn, D. Neves Micha, B. Bl\"asi, A. W. Bett, and F. Dimroth, Impact of Photon Recycling on GaAs Solar Cell Designs, IEEE J. Photovolt. \textbf{5}, 1636 (2015).

\bibitem{partanen_2017_sr}
M. Partanen, T. H\"ayrynen, and J. Oksanen, Interference-exact radiative transfer equation, Sci. Rep. \textbf{7}, 11534 (2017).

\bibitem{rte}
S. Chandrasekhar, \textit{Radiative Transfer} (Dover, New York, 1960).

\bibitem{partanen_2017}
M. Partanen, T. H\"ayrynen, J. Tulkki, and J. Oksanen, Quantized fluctuational electrodynamics for three-dimensional plasmonic structures, Phys. Rev. A \textbf{95}, 013848 (2017).

\bibitem{kivisaari_2018}
P. Kivisaari, M. Partanen, and J. Oksanen, Optical admittance method for light-matter interaction in lossy planar resonators, Phys. Rev. E \textbf{98}, 063304 (2018).

\bibitem{steiner_2013}
M. A. Steiner, J. F. Geisz, I. Garc\'ia, D. J. Friedman, A. Duda, and S. R. Kurtz, Optical enhancement of the open-circuit voltage in high quality GaAs solar cells, J. Appl. Phys. \textbf{113}, 123109 (2013).

\bibitem{wurfel_1982}
P W\"urfel, The chemical potential of radiation, J. Phys. C \textbf{15}, 3967 (1982).

\bibitem{heikkila_2009}
O. Heikkil\"a, J. Oksanen, and J. Tulkki, Ultimate limit and temperature dependency of light-emitting diode efficiency, J. Appl. Phys. \textbf{105}, 093119 (2009).

\bibitem{astm_2012}
ASTM G173-03(2012), Standard Tables for Reference Solar Spectral Irradiances: Direct Normal and Hemispherical on 37$^{\circ}$ Tilted Surface, ASTM International, West Conshohocken, PA, 2012, www.astm.org. Accessed August 15th, 2019.

\bibitem{niemeyer_2019}
M. Niemeyer, P. Kleinschmidt, A. W. Walker, L. E. Mundt, C. Timm, R. Lang, T. Hannappel, and D. Lackner, Measurement of the non-radiative minorityrecombination lifetime and the effectiveradiative recombination coefficient in GaAs, AIP Adv. \textbf{9}, 045034 (2019).

\bibitem{chen_2019}
H.-L. Chen, A. Cattoni, R. De L\'epinau, A. W. Walker, O. H\"ohn, D. Lackner, G. Siefer, M. Faustini, N. Vandamme, J. Goffard, B. Behaghel, C. Dupuis, N. Bardou, F. Dimroth, and S. Collin, A 19.9\%-efficient ultrathin  solar cell based on a 205-nm-thick GaAs absorber and a silver nanostructured back mirror, Nat. Energy \textbf{4}, 761 (2019).

\bibitem{yang_2014}
W. Yang, J. Becker, S. Liu, Y.-S. Kuo, J.-J. Li, B. Landini, K. Campman, and Y.-H. Zhang, Ultra-thin GaAs single-junction solar cells integrated with a reflective back scattering layer, J. Appl. Phys. \textbf{115}, 203105 (2014).

\bibitem{gai_2017}
B. Gai, Y. Sun, H. Lim, H. Chen, J. Faucher, M. J. Lee, and J. Yoon, Multilayer-Grown Ultrathin Nanostructured GaAs Solar Cells as a Cost-Competitive Materials Platform for III--V Photovoltaics, ACS Nano \textbf{11}, 992 (2017).

\bibitem{vaneerden_2020}
M. van Eerden, J. van Gastel, G. J. Bauhuis, P. Mulder, E. Vlieg, and J. J. Schermer, Observation and implications of the Franz-Keldysh effect in ultrathin GaAs solar cells, Prog. Photovolt.: Res. Appl. \textbf{28}, 779 (2020).

\bibitem{timo_2020}
G. Tim\`o, A. Martinelli, and L. C. Andreani, A new theoretical approach for the performance simulation of multijunction solar cells, Prog. Photovolt.: Res. Appl., \textbf{28}, 279 (2020).

\bibitem{xu_2019}
H. Xu, A. Delamarre, B. M. F. Yu Jeco, H. Johnson, K. Watanabe, Y. Okada, J.-F. Guillemoles, Y. Nakano, and M. Sugiyama, Current transport efficiency analysis of multijunction solar cells by luminescence imaging, Prog. Photovolt.: Res. Appl. \textbf{27}, 835 (2019).

\bibitem{sadi_2020}
T. Sadi, I. Radevici, and J. Oksanen, Thermophotonic cooling with light-emitting diodes, Nat. Photon. \textbf{14}, 205 (2020).

\bibitem{fan_2020}
D. Fan, T. Burger, S. McSherry, B. Lee, A. Lenert, and S. R. Forrest, Near-perfect photon utilization in an air-bridge thermophotovoltaic cell, Nature \textbf{586}, 237 (2020).

\bibitem{zhu_2019}
L. Zhu, A. Fiorino, D. Thompson, R. Mittapally, E. Meyhofer, and P. Reddy, Near-field photonic cooling through control of the chemical potential of photons, Nature \textbf{566}, 239 (2019).

\bibitem{wang_2011}
L. P. Wang, S. Basu, and Z. Zhang, Direct and Indirect Methods for Calculating Thermal Emission From Layered Structures With Nonuniform Temperatures, J. Heat Transfer \textbf{133}, 072701 (2011).

\bibitem{ren_2017}
X. Ren, Z. Wang, W. E. I. Sha, and W. C. H. Choy, Exploring the Way To Approach the Efficiency Limit of Perovskite Solar Cells by Drift-Diffusion Model, ACS Photon. \textbf{4}, 934 (2017).

\bibitem{vurgaftman_2001}
I. Vurgaftman and J. R. Meyer, Band parameters for III-V compound semiconductors and their alloys, J. Appl. Phys. \textbf{89}, 5815 (2001).

\bibitem{levinshtein_1996}
M. Levinshtein, S. Rumyantsev, and M. Shur, \textit{Handbook Series on Semiconductor Parameters: Volume 1: Si, Ge, C (Diamond), GaAs, GaP, GaSb, InAs, InP, InSb} (World Scientific, Singapore, 1996).

\bibitem{palik_gold}
D. W. Lynch and W. R. Hunter, in \textit{Handbook of Optical Constants of Solids}, edited by E. D. Palik (Academic Press, San Diego, 1998), pp. 286--295.

\bibitem{palik_1985}
E. D. Palik, in \textit{Handbook of Optical Constants of Solids}, edited by E. D. Palik (Academic Press, San Diego, 1998), pp. 429--443. 

\bibitem{palik_algaas}
O. J. Glembocki and K. Takarabe, in \textit{Handbook of Optical Constants of Solids}, edited by E. D. Palik (Academic Press, San Diego, 1998), pp. 513--558.

\bibitem{schubert_1995}
M. Schubert, V. Gottschalch, C. M. Herzinger, H. Yao, P. G. Snyder, and J. A. Woollam, Optical constants of Ga$_x$In$_{1-x}$P lattice matched to GaAs, J. Appl. Phys. \textbf{77}, 3416 (1995).

\bibitem{linnik_2002}
M. Linnik and A. Christou, Calculations of optical properties for quaternary III-V semiconductor alloys in the transparent region and above (0.2--4.0 eV), Phys. B: Cond. Matt. \textbf{318}, 140 (2002).

\bibitem{ozaki_1993}
S. Ozaki and S. Adachi, Optical constants of cubic ZnS, Jpn. J. Appl. Phys. \textbf{32}, 5008 (1993).

\bibitem{rodriguez_2017}
L. V. Rodr\'iguez-de Marcos, J. I. Larruquert, J. A. M\'endez, J. A. Azn\'arez, Self-consistent optical constants of MgF$_2$, LaF$_3$, and CeF$_3$ films, Opt. Mater. Express \textbf{7}, 989 (2017).

\bibitem{polanskiy}
M. N. Polyanskiy, "Refractive index database," https://refractiveindex.info. Accessed August 15th, 2019.

\end{thebibliography}
\end{document}